\documentstyle[preprint,aps]{revtex}
\begin{document}
\draft
\begin{title}
  {\Large \bf Layer dynamics of a freely standing smectic-A film
  }
\end{title}

\author{Hsuan-Yi Chen and  David Jasnow}

\address{Department of Physics and Astronomy, University of Pittsburgh, 
  Pittsburgh, PA 15260, U.S.A.}

\date{\today}
\maketitle
\begin{abstract}
   We study the hydrodynamics of a freely-standing smectic-A film
in the isothermal, incompressible limit theoretically by analyzing 
the linearized 
hydrodynamic equations of motion with proper boundary conditions. 
The dynamic properties for the system can be obtained
from the response functions for the free surfaces.
Permeation is included and its importance near the free surfaces 
is discussed. The hydrodynamic mode structure for the dynamics 
of the system is compared with that of bulk systems.   
We show that to describe the dynamic correlation functions for the system, 
in general, it is 
necessary to consider the smectic layer displacement $u$ and 
the velocity 
normal to the layers, $v_z$, together.
Finally, our analysis also provides a basis for the theoretical study 
of the off-equilibrium dynamics of freely-standing smectic-A films.
\\[1ex]
 Mar 15, 1999; printed \today

\end{abstract}

\pacs{61.30.Cz, 68.15.+e, 83.70.Jr}
\narrowtext

\section{INTRODUCTION}
A three dimensional smectic-A phase is at its lower critical 
dimension: the  smectic layer displacement fluctuations grow 
logarithmically with the size of the system, and the divergence 
is sufficiently weak 
that finite size effects stablize laboratory samples \cite{DeGennes1}.  

Smectic liquid crystals also form freely standing films with a 
controlled size and high degree of uniformity.  Due to the 
presence of two free surfaces
and finite thickness, experimental and and theoretical studies
on the equilibrium \cite{Mol,Holyst,Fournier,Toner} , 
and off-equilibrium \cite{Wu}
properties of freely standing smectic-A films show drastically 
different properties from bulk systems.  Hence freely standing 
smectic-A films are excellent materials to study the effect of 
reduced dimensionality and surface effects.

In addition to the static properties, the dynamics of a freely
standing smectic-A film have been studied in the ``low'' frequency
($\omega \ll 50$kHz) and ``long'' wavelength ($\lambda \gg 0.1$ cm)
regime, where it behaves like a simple elastic membrane \cite{Miyano}.
The dynamic correlation functions have also been considered in the
strongly overdamped regime, \cite{Toner} both theoretically and
experimentally.  Notice, however, that these experiments were 
performed in a region where the effects of bulk elasticity are 
negligible; further experiments are needed to probe the dynamical 
crossover behavior of the system.

In this article we introduce a continuum theory for the dynamics of a
smectic-A film in the isothermal, incompressible limit based on the
linear hydrodynamic theory constructed by Martin {\it et al}
\cite{Martin}.  Our study provides a finite-thickness counterpart to
the hydrodynamic theory for bulk smectic-A systems.  The following two
questions are addressed and answered in our theory: (i) Permeation
plays a role near the free surfaces \cite{Orsay}, but how does it
modify the smectic layer dynamics near the boundaries?  (ii) The
smectic layer displacement $u$, and the velocity of the liquid crystal
normal to the layers, $v_z$, are coupled hydrodynamic variables, but
how do the hydrodynamic modes for $u$ and $v_z$ for a film compare to
the linear theory for a bulk smectic-A \cite{DeGennes1,Martin} ?  In
the following sections we will show that the dynamic properties of the
smectic layers in a freely standing film can be extracted from the
response functions of the free surfaces.  Since the layer displacement
for a smectic-A film has to satisfy certain boundary conditions on the
free surfaces, the normal modes for the dynamics of the layers depend
on both surface and bulk properties.  We find that among all the
normal modes, two are most strongly affected by the surface tension
and film thickness. The decay rates for all other normal modes show
the interplay of bulk-like behavior and effects from the
surfaces.  When the thickness of the film is small, the dominant
thermal fluctuations of the layers come from these two modes, and the
system can behave similarly to an ordinary fluid film.  When the
thickness becomes large, the other normal modes have sufficiently
large contributions to the layer fluctuations, and the internal layer
structure of the system can be detected experimentally.  
The present analysis reveals how the dynamics of a finite thickness 
smectic-A system depends on surface tension, viscosity, and also bulk
elasticity.  We also provide a theoretical picture for the crossover
behavior of a freely standing smectic-A film from a thin film, which
depends only on the surface properties, to a three-dimensional system
with layer structure.  Furthermore, this theory is also a starting
point for theoretical calculations for the off-equilibrium dynamics of
a freely standing smectic-A film.

    The remainder of this paper is organized as follows.  Section II
contains a review of the linear elastic and hydrodynamic theory for
a bulk smectic-A in the isothermal, incompressible limit.  After a brief
review of the linear response functions of
the surfaces of a smectic-A film, we extract the smectic 
layer dynamics and compare the bulk and film hydrodynamics in
Section III.  The crossover of smectic dynamics from film-like
behavior to bulk-like behavior is discussed in Section IV.  In 
Section V
we summarize results and add concluding remarks. 
Some additional details on
permeation near the surfaces and the calculation of the dynamic
correlation functions are provided in Appendices A and B, respectively.

\section{Elasticity and hydrodynamics of bulk smectic-A phase}
\label{sec:bulk}
The symmetry broken direction for a smectic-A system is chosen to be
the $z$-direction.  
Assume that temperature and density fluctuations occur
on time scales much faster than the characteristic times under 
consideration. In this limit we have an isothermal, incompressible 
system, and the elastic free energy of the smectic layer distortion 
for a bulk smectic-A
in thermal equilibrium is provided to leading order by \cite{DeGennes1}
\begin{eqnarray}
F = \frac{1}{2} \ \int dV 
        \left\{B \left(\frac{\partial u}{\partial z}\right)^2 + 
        K_1 \left(\frac{\partial ^2 u}{\partial x^2} 
        + \frac{\partial ^2 u}{\partial y^2}\right)^2 \right\} 
 \end{eqnarray}
where $u({\bf r},t)$ is the layer displacement, 
${\bf r}=(x,y,z)$, and $B$ and $K_1$ are, respectively, 
the layer compression and undulation elastic moduli.  
The volume integral is performed over the bulk. In this article we
consider exclusively a thermotropic smectic-A, for which typically
$B \sim 2.5 \times 10^7 \ {\rm dyn/cm^2}$, $K_1 \sim 10^{-6}$ dyn.  
The characteristic
length  $\lambda \equiv \sqrt{K_1/B}$ is typically on the order of the
layer spacing ($ \sim 10 ^{-7}$cm)\cite{DeGennes1}.  It has the following
physical interpretation: in bulk systems
the dominant thermal fluctuations of the layers with momentum {\bf q} 
are those with $q_z \sim \lambda q_{\perp}^2$, where $q_{\perp}$ is the 
magnitude of the projection of {\bf q} on the $xy$-plane.  
When $q_{\perp} \ll \lambda^{-1},$ the dominant layer fluctuations come from 
$q_z \ll q_{\perp}$.

The dynamics of a bulk smectic-A have been studied 
thoroughly in, for example, Ref.~\cite{Martin}, which we follow. 
In the absence of 
topological defects, under constant temperature and assuming 
incompressibility, to linear order the system satisfies the equations 
of motion 
\begin{eqnarray}
    \rho \frac{\partial v_i}{\partial t}=-\partial_i p
+ \partial_j  \sigma' _{ij} + h  \ \delta _{iz}
\label{NS}
\end{eqnarray}
and 
\begin{eqnarray}
    \frac{\partial u}{\partial t}=v_z+\zeta_p \ h \ \ ,
\label{dudt}
\end{eqnarray}
where $\rho$ is the density (typically $\rho \sim 1 \ {\rm g/cm^3}$),
$v_i$ is the $i$th component of the velocity field, and the pressure, $p$, 
is actually a Lagrange multiplier for the incompressibility condition.
The viscous stress tensor is denoted by  $\sigma'$ \cite{Martin}, $\zeta_p$ 
is the permeation
constant, and the molecular field $h$ is defined by 
\begin{eqnarray}
   h \equiv \partial_i \left( \frac{\delta F}{\delta \partial_i u}\right) \ .
\label{molfield}
\end{eqnarray}
In Eqs.~(\ref{NS}) and (\ref{molfield}) we sum on repeated indices, and 
$\partial_j  = \partial / \partial x_j.$
 For layer displacements of the form 
\begin{eqnarray}
 u({\bf r},t) =
 u({\bf q},\omega) e^{i({\bf q}_{\perp} \cdot {\bf r}_{\perp}
                               +q_z z) -i\omega t} 
\label{u(q)}                               
\end{eqnarray}
with $q_z \ll q_{\perp}$, permeation may be neglected \cite{DeGennes1,Martin}, 
and the relation between $\omega$ and {\bf q} is given by
\begin{eqnarray}
  i \omega
= \frac{\eta_3 q_{\perp}^2}{2\rho} \times
     \left(1 \pm \sqrt{
             1-\frac{4\rho}{\eta_3^2 q_{\perp}^4} (Bq_z^2+K_1q_{\perp}^4) 
                       }
     \right)
\equiv i \omega_{\pm}(q_{\perp},q_z) \ ,     
\label{iomega}     
\end{eqnarray}
where $\eta_3$ (typically $\eta _3 \sim 1 $p) is one of the five viscosity 
coefficients of a smectic-A \cite{Martin}.
The fact that $u({\bf r},t)$ couples to $v_z({\bf r},t)$ in the equations 
of motion means that, in general, a hydrodynamic normal mode described
by Eq.~(\ref{iomega}) is a linear combination of these two variables.
There are two other normal modes for a given ${\bf q}$, but they correspond
to the diffusive modes of $v_x$ and $v_y$, and they are decoupled from
the motion of the smectic layers~\cite{Lubensky}.  
We will not discuss these two other modes in the rest of this section.
In the limit $4\rho(Bq_z^2+K_1q_{\perp}^4)/\eta_3^2 q_{\perp}^4\ll 1$, the
viscous loss of the system is large, and the motion of the layers
is strongly overdamped. Then $i \omega$ is real, and the two decay 
rates are given by 
\begin{eqnarray}
 \frac{1}{\tau _1} &=& \frac{\eta_3 q_{\perp}^2}{\rho}    \nonumber \\
 \frac{1}{\tau _2} &=&\frac{Bq_z^2+K_1q_{\perp}^4}{\eta_3 q_{\perp}^2},
\label{tau2}
\end{eqnarray}
where $1/\tau _1 \gg 1/\tau _2$.  We can identify $1/\tau_1$ as the decay rate
of the diffusive mode associated with $v_z$ and $1/\tau_2$ as the decay rate
for the smectic layer undulations. Since we are interested in the 
low frequency behavior of the system, it is appropriate to neglect the 
effect of $\tau_1$.
In the opposite limit, \ $4\rho(Bq_z^2+K_1q_{\perp}^4)/\eta_3^2
q_{\perp}^4\gg 1$,  
the elastic restoring force is much stronger than the viscous loss, and 
the system performs underdamped oscillations, i.e., $i \omega$ becomes
complex.   The decay rate is simply
\ $\eta_3 q_{\perp}^2/2\rho$,  and the oscillation frequency is 
approximately \ $\sqrt{(Bq_z^2+K_1q_{\perp}^4)/\rho}$ \ , which is the 
frequency of ``second sound''\cite{DeGennes1}.  
In this case, then, there are two propagating hydrodynamic modes 
associated with each {\bf q}.
Notice that when $q_z \sim \lambda q_{\perp}^2$, the system is strongly 
damped as long as $\rho K_1 / \eta_3^2 \ll 1$;
for typical materials $\rho K_1 / \eta_3^2 \sim 10^{-6}$, so that 
the dominant
thermal fluctuations of the layers are overdamped.

\section{Hydrodynamic normal modes of a freely standing smectic-A film}
\label{sec:film}
Let us now turn to the case of freely standing smectic-A films. 
In the absence of external fields, the smectic layers are always aligned
and parallel to both free surfaces.  We consider a 
film which extends from $z=d/2$ to $z=-d/2$ in the vertical
direction and is of infinite extent in the $x$ and $y$-directions. 
The geometry of the system is shown in Fig.~1, which shows  
$\zeta^{+(-)}$ as the displacement of the upper (lower) surface from 
its equilibrium position. In this section we first briefly review 
previous results \cite{HY-D}
on the derivation of  the linear response function 
for the equilibrium surface fluctuations of a freely standing smectic-A film
in the low frequency, long wavelength regime 
\cite{l-l limit,com.HY-D}, then extend these results to the structure of
hydrodynamic normal modes of the system.

  Suppose the system were perturbed by external forces with fixed in-plane
wave vector ${\bf q}_{\perp}$ and frequency $\omega$ on both upper and lower 
surfaces, i.e, 
\begin{eqnarray}
  & P_{ext}^+&({\bf r}_{\perp},t)
   =P_{ext}^+({\bf q}_{\perp},\omega) \ 
    e^{i {\bf q}_{\perp} \cdot {\bf r}_{\perp} -i \omega t} \nonumber 
     \\
  & P_{ext}^-&({\bf r}_{\perp},t) =P_{ext}^-({\bf q}_{\perp},\omega) \
    e^{i {\bf q}_{\perp} \cdot {\bf r}_{\perp} -i \omega t} \ .
\end{eqnarray}
We look for the linear response functions of the surfaces in the regime
of weak external forces.
On the free surfaces the system has to satisfy
the following boundary conditions:
\begin{description}
\item[bc 1]
  The velocities of the free surfaces are the same as the normal 
component of 
the liquid crystal velocity on the surfaces.  
\item[bc 2]
         For free surfaces the normal component of the 
         permeation force should vanish.~\cite{HY-D} 
\item[bc 3] 
         The force acting on the system is continuous across the free 
         surfaces.
\end{description}
The boundary conditions lead to the following equations for the system,
\begin{eqnarray}
   \frac{\partial \zeta ^{+(-)}}{\partial t} 
 = \left[ v_z \right]_{z=\pm d/2}.
\label{zeta-u}
\end{eqnarray}
\begin{eqnarray}
  \left[ \frac{\partial u}{\partial z} \right]_{z=\pm d/2} =0 \ . 
\label{dudz}  
\end{eqnarray}
\begin{eqnarray}
  \left[ \partial _i v_z + \partial _z v_i \right]_{z=\pm d/2} = 0 
  \ , \ \ \  i=x, \ y \ , \\
\label{dvdz}  
\left[ \pm \left( -p + 2 \eta_3 \frac{\partial v_z}{\partial z}
                   + B \frac{\partial u}{\partial z} 
             \right)   
     - \alpha \nabla _{\perp} ^2 \zeta ^{\pm} - P_{ext}^{\pm}
\right]_{z=\pm d/2} =0  \ ,
\end{eqnarray}
where $\alpha$ is the surface tension 
(typically $\alpha \sim 30 \ {\rm dyn/cm}$).

Solving the equations of motion, {\it i.e.}, Eqs.~(\ref{NS}), and
(\ref{dudt}), with these boundary conditions, one finds that, 
in the presence
of the driving forces with given q$_{\perp}$ and $\omega$, 
there are four different $q_z$'s.
One of them is associated with the diffusive motion for $v_x$ and $v_y$,
and is decoupled from the motion of $u$ and $v_z$; hence it is 
irrelevant
for the calculation of surface response functions (see bc 1).
Another two values of $q_z$ are associated with permeation, 
and these contributions to the layer displacement are small
(see Appendix A). 
The last of these $q_z$-values dominates the 
contribution to the smectic layer displacement; hence $u$ 
can be approximated by
\begin{eqnarray}
u({\bf r},t) &=& \mbox{ } u_S({\bf q}_{\perp},\omega) \ 
                        {\rm cos}\left[q_z(q_{\perp},\omega) z\right]    
                  \times e^{i{\bf q_{\perp}\cdot r_{\perp}} -i\omega t
                          } \nonumber  \\
             & & + u_A({\bf q}_{\perp},\omega) \ 
                        {\rm sin}\left[q_z(q_{\perp},\omega) z\right]       
                 \times e^{i{\bf q_{\perp}\cdot r_{\perp}} -i\omega t
                          }   \ \ .
\label{u-solution}
\end{eqnarray}               
Using $\omega_{\pm}(q_{\perp},q_z)$ as defined in Eq.(\ref{iomega}),
$q_z(q_{\perp},\omega)$ is determined by the solution of the
following with Re$q_z >0$:
\begin{eqnarray}
  i \omega = i \omega _{\pm}(q_{\perp},q_z),
\end{eqnarray}
where $|q_z| \ll q_{\perp}$, and, in general, $q_z$ is complex
under the conditions considered~\cite{l-l limit}.
Recall that $\omega$ and $q_{\perp}$ describe
the driving forces and $q_z$ is not independent.
This differs from the interpretation of the previous section in which the 
bulk hydrodynamics of a smectic-A was considered in the absence of 
external driving forces.

When $P_{ext}^+=P_{ext}^-$ ($P_{ext}^+=-P_{ext}^-$), 
the dynamics of the system are
symmetric (antisymmetric) under $z\rightarrow -z$, and
the layer displacement is described by $u_S$ ($u_A$) alone.
We define the response functions for both
symmetric and anti-symmetric surface motion as
\begin{eqnarray}
  \zeta^S (q_{\perp},w)
= X^S(q_{\perp},w) \left[  P_{ext}^+(q_{\perp},w)+P_{ext}^-(q_{\perp},w,) 
                            \right] \ ,
  \nonumber \\
  \zeta^A (q_{\perp},w)
= X^A(q_{\perp},w) \left[  P_{ext}^+(q_{\perp},w)-P_{ext}^-(q_{\perp},w,) 
                            \right] \ ,
\label{res}   
\end{eqnarray}
where
\begin{eqnarray}
\zeta ^{S(A)}= \frac{1}{2} \left(\zeta ^+ \pm \zeta ^- \right).
\end{eqnarray}
A calculation from the continuum hydrodynamics leads to the following
expressions for  the response functions~\cite{HY-D},
\begin{eqnarray}
X^{S} = \frac{1}{2} \ \ 
        \frac{1}{\alpha q_{\perp}^2 - B \ q_z(q_{\perp},\omega) \ 
        {\rm tan}\left(d \ q_z(q_{\perp},\omega)/2 \right) }  \ ,
\label{undu} \\                       
X^{A} =\frac{1}{2} \ \ 
       \frac{1}{\alpha q_{\perp}^2 + B \ q_z(q_{\perp},\omega) \ 
       {\rm cot} \left(d \ q_z(q_{\perp},\omega)/2 \right) }  \  . 
\label{per}                        
\end{eqnarray}       

It appears that the permeation processes have no contribution
to dynamics of the surfaces in the regime \cite{l-l limit} 
where our analysis is done.  However, the solution for
$u({\bf r},t)$ in Eq.~(\ref{u-solution}) does not strictly satisfy 
bc 2.
This means that the contribution from permeation enables the system 
to satisfy the boundary conditions~\cite{Orsay} but otherwise has little
significance.  The role of permeation in the dynamics of the system 
is discussed in further detail in Appendix A.

We have derived the response functions
$X^{S(A)}$ as functions of in-plane wave vector $q_{\perp}$ and 
frequency $\omega$ of the applied external forces.  We can also 
define $\omega$ on the complex plane; 
then the poles of the response
functions provide the frequencies of the surface fluctuations in the 
absence of
driving forces~\cite{Lubensky}.  In linear theory these poles also reveal
the frequencies of layer displacement.  
Each pole in the complex $\omega$-plane 
corresponds to a hydrodynamic normal mode associated with
the layer displacement.  Since we are not interested in the 
dynamics of $v_x$, and $v_y$, which decouple from the
dynamics of $u$ and $v_z$, these poles provide the information
of interest.
Two classes of normal modes 
with opposite symmetry properties   
can be found.  One class of normal modes is symmetric under 
$z \rightarrow -z$; the frequencies are the roots of,
\begin{eqnarray}
  \frac{B q_z(q_{\perp}, \omega)}{\alpha q_{\perp}^2}  
= {\rm cot}\left(\frac{q_z(q_{\perp}, \omega) d}{2}\right) \ .
\label{undulpole}
\end{eqnarray}
The other class of normal modes is anti-symmetric under 
$z \rightarrow -z,$ and
their frequencies are determined by
\begin{eqnarray}
  \frac{B q_z(q_{\perp}, \omega)}{\alpha q_{\perp}^2} 
= -{\rm tan}\left(\frac{q_z(q_{\perp}, \omega) d}{2}\right) \ .
\label{peripole}
\end{eqnarray}
Actually, we can go back and treat the above two equations as equations 
for $q_z$; we can find the $z$ dependence of the layer displacement 
for the normal modes from these equations.  
The form of the layer displacement in Eq.~(\ref{u-solution}) suggests
that, for given ${\bf q}_{\perp},$ the layer displacement for an 
normal mode is an eigenfunction of the differential equation
\begin{eqnarray}
  \frac{\partial ^2 \psi({\bf q_{\perp}},z)}{\partial z^2} 
= -q_z ^2 \ \psi({\bf q_{\perp}},z)  \ 
\end{eqnarray}
with boundary condition
\begin{eqnarray}
\lefteqn{ {\bf bc 2'} }  \hspace{0.85in}
  \left[ B \ \frac{\partial \psi({\bf q_{\perp}},z)}{\partial z} 
         \pm \alpha \ q_{\perp}^2 \ \psi({\bf q_{\perp}},z) 
  \right]_{z=\pm d/2} 
  = 0 \ ,
\label{nmode}
\end{eqnarray}
It is straightforward to show that these eigenfunctions form a 
complete basis for the interval $-d/2 < z < d/2$.  
The apparent contradiction between bc 2' and
bc 2 is discussed in Appendix A.

In general $q_z$ in Eqs.~(\ref{undulpole}) and (\ref{peripole}) cannot be
determined analytically.  However, the $q_z$'s are all real, 
and we can choose all
of them to be positive.
The graphical solutions for the $q_z$'s are shown 
in Fig.~2.  We label the $q_z$'s for the normal modes 
as $q_z^{(n)}(q_{\perp}), 
 \ n=0,1,2, ....$ with $
q_z^{(0)}(q_{\perp}) <  q_z^{(1)}(q_{\perp}) <  q_z^{(2)}(q_{\perp}) 
<  ....$,
and note there are two normal modes associated with each $q_z^{(n)}$. 
The layer displacement for normal modes is even (odd) under 
$z \rightarrow -z$
if $n$ is even (odd).  The layer displacement for the normal modes with 
$n=0$
is even under $z \rightarrow -z$. When both modes are underdamped,
they form a pair of propagating waves, which in bulk systems corresponds to 
second sound \cite{DeGennes1}.  When they are overdamped,
they are two diffusive modes.  In the {\it strongly} overdamped limit, 
the characteristic time scales are well separated; the fast mode 
corresponds
to the diffusive motion of $v_z$, and the slow mode corresponds to 
the layer undulation mode in bulk systems.

To compare with the behavior of a bulk smectic-A, recall
that translational invariance of the bulk system allows one to label
modes by a $d$-dimensional wave vector.
For a freely
standing smectic-A film, the boundary conditions and the equations of motion
select a set of normal modes for the dynamics of the system.  We can
express the normalized time-independent part of  the normal modes 
in the following way,
\begin{eqnarray}
 \psi({\bf q}_{\perp},n;z) 
= N_S(q_{\perp},n) \ {\rm cos}\left[q_z^{(n)}(q_{\perp}) z\right]  \ ,
\label{normals}        
\end{eqnarray}
for $n=0,2,4,...$,
and
\begin{eqnarray}
 \psi({\bf q}_{\perp},n;z) 
=N_A(q_{\perp},n) \ {\rm sin}\left[q_z^{(n)}(q_{\perp}) z\right]  \ ,
\label{normala}       
\end{eqnarray}
for $n=1,3,5,...$ , where 
\begin{eqnarray}
   N_S(q_{\perp},n) 
 = \sqrt{\frac{2}{d}} \ 
   \left( 
        1+ \frac{{\rm sin}(q_z^{(n)} d)}{q_z^{(n)} d} 
   \right)^{-1/2} \ , 
\nonumber   \\
   N_A(q_{\perp},n) 
 = \sqrt{\frac{2}{d}} \ 
   \left( 1- \frac{{\rm sin}(q_z^{(n)} d)}{q_z^{(n)} d} \right)^{-1/2} \ . 
\end{eqnarray}
These normal modes satisfy the orthonormality condition
\begin{eqnarray}
 \int _{-d/2} ^{d/2} dz \ 
     \psi({\bf q}_{\perp},n;z) \psi({\bf q}_{\perp},m;z)
=\delta _{mn} \ .
\end{eqnarray}     
Hence, these modes can be used as a basis to expand any 
$u({\bf q_{\perp}},z,t)$ as, 
\begin{eqnarray}
   u({\bf q}_{\perp},z,t) = \sum _n u({\bf q}_{\perp}, n,t) 
                                  \times \psi ({\bf q}_{\perp},n;z) \ .
\end{eqnarray}                                  
From these relations one finds the dynamic correlation function for $u$. 
As discussed in Appendix B, the general form for the dynamic
correlation function is
\begin{eqnarray}
& & c(q_{\perp},n,t)   \nonumber  \\
&\equiv& < u(q_{\perp},n, t) \  u(-q_{\perp},m, 0) >   \nonumber  \\
&=& \delta_{nm}  \
   \frac{i\omega_+ e^{-i\omega_- t} - i\omega_- e^{-i\omega_+ t}}
        {i\omega_+-i\omega_-} 
   \times
   \frac{k_B T}{B {q_z^{(n)}}^2 + K_1 q_{\perp}^4} \ ,    
\end{eqnarray}                          
where $\omega_{\pm} = \omega_{\pm}(q_{\perp},q_z^{(n)})$ is related to 
$q_{\perp}$ and $q_z^{(n)}$ through Eq.~(\ref{iomega}).  In the
{\it underdamped} case, $\omega_{\pm}$ 
are complex and $\omega_- = \omega_+^*$.  
The dynamic correlation function has a 
damping rate $\eta_3 q_{\perp}^2/2 \rho$ and oscillates with frequency 
$\sqrt{-(\eta_3 q_{\perp}^2 / 2 \rho)^2 + (B{q_z^{(n)}}^2 
+ K_1 q_{\perp}^4)
/\rho}$.  In the {\it overdamped} case, two decay rates exist, and the 
fast mode can be neglected only when the time scales are well separated.  
In general,  when an experiment is performed outside the strongly 
overdamped regime, two characteristic time scales should be detected.  
For weakly overdamped motion, these are the two decay rates, 
while for underdamped 
motion, these are the real and imaginary parts of $\omega_{\pm}$.
 
In summary, we find that in common with analysis of a liquid film, the
normal modes are symmetric/anti-symmetric under $z \rightarrow -z$.  
They also
are labeled by $({\bf q_{\perp}}, n)$, and the layer displacement 
can be treated as eigenfunctions of a differential equation.
The frequencies for those normal modes can be obtained from the poles of
the {\it surface response functions} in the complex $\omega$-plane.
As will become clear in the next section, the difference between
those $q_z^{(n)}$'s which correspond to the normal modes
for a finite thickness smectic-A film, and the set of numbers 
$n \pi/d$, $n=1,2,3,...$ provides a measure of the degree to which finite 
thickness and surface tension change the dynamic properties of the layer 
displacement.  Experimentally, these effects are 
typically obtained by measuring 
the frequencies and the magnitude of the autocorrelation function of 
the layer 
displacement.  Layer dynamics well be discussed further in the following 
section.

\section{Crossover of layer dynamics for smectic-A films}
 There are two special regimes where the $q_z^{(n)}$'s can 
be evaluated approximately, and the physics of the system is 
also of interest.  We call these the {\it thin film} and {\it thick film}
regimes, respectively.  We will show that
the pair of normal modes with $q_z^{(0)}$ is special since they strongly
depend on the surface properties and show drastically different behavior 
from the thin film to thick film regime.

    When the dimensionless measure of film thickness 
$\delta \equiv \alpha q_{\perp}^2 d/B \ll 1$, the solution for $q_z$ 
can be approximated by 
\begin{eqnarray}
    q_z^{(0)} = \frac{ \sqrt{2 \delta} }{d}
              = \sqrt{\frac{2\alpha}{Bd}} \ q_{\perp}
\label{thin-first}
\end{eqnarray}
and
\begin{eqnarray}
q_z^{(n)}= \frac{n\pi}{d}
  	      \left( 1 + \frac{2 \delta}{(n\pi)^2}\right) 
  	 = \left( \frac{n\pi}{d} + 
           \frac{2\alpha q_{\perp}^2}{n\pi B} \right)     \ ,
          \ \ \ n=1,2,3,...
\label{thin-others}
\end{eqnarray}
For $n >0$ the modes are spaced approximately by $\pi/d$.  
This is the {\it thin film} regime ($\delta \ll 1$).

   The equilibrium thermal fluctuations of the smectic layers in this limit
are dominated by surface effects.  This can be illustrated by 
considering the pair correlation function for the layer displacement.  
From Appendix A, in thermal equilibrium, for a smectic-A film with given
${\bf q}_{\perp}$, the contribution to the pair correlaton function from
the 
modes with $q_z^{(0)}$ is
\begin{eqnarray}
 \langle |u({\bf q}_{\perp},0)|^2 \rangle
=\frac{k_B T}{2\alpha q_{\perp}^2/d + K_1 q_{\perp}^4}.
\label{0-flu}
\end{eqnarray}
Furthermore, when 
$K_1 q_{\perp}^2 d/ \alpha \ll 1$, the contribution from layer 
elasticity is negligible, and the thermal fluctuations of the smectic layers
for these normal modes are independent of the bulk elastic 
coefficients~\cite{Toner}.
For a typical material, $B \sim 2.5 \times 10^7 \ {\rm dyn/cm^2}$, 
$K_1 \sim 10^{-6}$ dyn, $\alpha \sim 30 \ {\rm  dyn/cm}$, and
the thin film regime corresponds to 
$q_{\perp}^2 d \ll B/\alpha \sim 10^6 \ {\rm cm}^{-1}$.
The layer fluctuation is dominated by the surface tension term when
$q_{\perp}^2 d  \ll \alpha / K_1 \sim 3 \times 10^7 \ {\rm cm}^{-1}$, 
{\it i.e.}, in the thin film regime, layer fluctuations due to these 
two normal
modes are typically dominated by the surface tension term.  In the 
following
we will simply drop the contribution from the $K_1$ elasticity term in the 
thin film regime.

The total contribution from the fluctuations of the other normal modes for
fixed $q_{\perp}$ is approximately
\begin{eqnarray}
& &\sum_{n=1}
 \frac{k_B T}
   {B \left( n\pi/d + 2\alpha q_{\perp}^2/n\pi B \right)^2} \nonumber \\
&\sim &  \frac{k_B T d^2}{B} \ .
\end{eqnarray}
This contribution is small compared
to the contribution from the normal modes with $q_z^{(0)}$ in thin film limit. 
As a result, we find that a freely standing smectic-A film can be
treated as an 
ordinary fluid film with surface tension $\alpha$ and thickness $d$ in this 
limit.\cite{Toner}  The internal structure of the layers is not important.
Comparing with bulk smectic-A systems where
the dominant layer fluctuations for given $q_{\perp}$
come from $q_z \sim \lambda q_{\perp}^2$ , for a thin film the dominant layer 
fluctuations come from only two normal modes, and it is in general not true
that $q_z^{(0)} \sim \lambda q_{\perp}^2$.  

From Eq.~(\ref{iomega}) the normal mode frequencies
$\omega_{\pm}(q_{\perp},q_z^{(0)})$ are given by 
\begin{eqnarray}
 i\omega_{\pm}(q_{\perp},q_z^{(0)})
= \frac{\eta_3 q_{\perp}^2}{2\rho} \times
   \left( 1 \pm \sqrt{1-\left( \frac{q_c}{q_{\perp}} \right)^2 } 
   \right)  \ ,
\end{eqnarray} 
where $q_c=\sqrt{8\rho\alpha/\eta_3^2 d}$ is a characteristic
wave number of the film.
In the case of short wavelength fluctuations with  
$q_{\perp} \gg q_c$,
viscous loss is strong compared to elastic effects, 
and the overdamped decay rates are
\begin{eqnarray}
 i\omega_+^{(0)}(q_{\perp}) = \frac{\eta_3 q_{\perp}^2}{\rho} \ ,
 \nonumber \\
 i\omega_-^{(0)}(q_{\perp}) = \frac{2\alpha}{\eta_3 d} \ .
\end{eqnarray}
Since we are interested in slow motion of the system, 
$i\omega_-^{(0)}(q_{\perp})$
provides the characteristic time for the system.
Previous experimental studies of dynamic correlation functions were 
performed in this regime \cite{Toner}.
Notice that in this regime the decay rate is independent of the 
magnitude of the
in-plane wave vector ${\bf q}_{\perp}$ and bulk elastic constants.  
In general, experiments which probe the fluctuations of the film 
with smaller
in-plane wave vector should be able to identify two decay rates.  
This is especially so when the motion of the system is close to 
being critically damped,
as can be seen from Fig.~3.  
For long wavelength $q_{\perp} \ll q_c$ 
the film oscillates
with frequency $\sqrt{2\alpha/\rho d} \ q_{\perp}$ and damping rate
$\eta_3 q_{\perp}^2/2\rho$. Again, both time scales in this range of 
$q_{\perp}$ are independent of $B$ and $K_1$.  
As $q_{\perp} \rightarrow 0$, the damping 
is small; this is the regime in which the long wavelength theory and 
experiments\cite{Miyano} have been performed.

   We have shown that the physical properties for the system in the
{\it thin film} regime are basically the same as a two dimensional
object, {\it i.e.}, a {\it simple fluid film}.  The dynamics of the
system show a competition between the elasticity due to {\it surface
tension} and viscous loss characterized by the coefficient $\eta_3$.
The underdamped motion for a smectic-A {\it film} should not be
confused with ``second sound'' for smectic layers, because for the
thin film the layer compression is negligible.  In this regime, the
underdamped motion is simply the vibration of a simple fluid film with
surface tension $\alpha$.  Also notice that although the magnitude of
$q_z^{(0)}$ depends on the elastic constant $B$, this dependence does 
not show up in the dynamical
correlation function for the hydrodynamic normal modes, as can be 
seen in Eq.~(\ref{0-flu}), which is the
same as that for the surface fluctuation.

To probe the crossover behavior of the film, experiments have to be
performed outside the thin film regime so that the effect of the bulk
elasticity can be revealed.  This can be done by increasing the film
thickness or adjusting the in-plane wave vector.  When $\alpha
q_{\perp}^2 d /B$ increases to order unity, the layer structure of the
system starts to play a non-negligible role in the system.  First of
all, this is reflected in the fact that the approximation for
$q_z^{(0)}$ in Eq.~(\ref{thin-first}) is poor, and the higher order
terms contribute. Then the $n=0$ modes start to show the influence of
bulk elasticity.  Second, the layer fluctuations of all other
normal modes become comparable to the layer fluctuations for the $n=0$
modes.  Hence, the dynamics of the layers for fixed $q_{\perp}$ are no
longer simply controlled by $\omega_+(q_{\perp},q_z^{(0)})$ and
$\omega_-(q_{\perp},q_z^{(0)})$, and this means that there are more
relevant degrees of freedom in the system than the displacement of the
two surfaces.  The first sign of this crossover behavior can be
detected by exciting the $n=1$ modes.  As pointed out in the study of
the surface dynamics of freely standing smectic-A 
films~\cite{HY-D,peristaltic}, this pair of modes is similar 
to the
so-called {\it peristaltic} mode for a soap film.  However, the
characteristic time scales of these modes strongly depend on the
elasticity of the smectic layers; hence, the dynamic properties of
these modes actually reveal the difference between a smectic film and
a soap film.  Possible future surface light scattering experiments for
probing these normal modes are also discussed in Ref.~\cite{HY-D}.

For $\delta \gg 1$, $q_z^{(0)}$ increases and $q_z^{(0)}
d \sim \pi$; the $q_z$'s for all normal modes can be approximated by a
single expression,
\begin{eqnarray}
 q_z^{(n)} = \frac{(n+1)\pi}{d} \left( 1-\frac{2}{\delta} \right)
 	   = \frac{(n+1)\pi}{d} 
               \left( 1-\frac{2B}{\alpha q_{\perp}^2 d} \right), \ \ \ 
    n=0,1,2,3,...
\end{eqnarray}    
From Eqs.~(\ref{normals}) and  (\ref{normala}) we find that the 
layer displacement
for the normal modes on both free surfaces becomes vanishingly 
small as 
$\delta^{-1} \rightarrow 0$.  This means that the 
layer displacement
is not affected by the surface tension in this limit, 
and the behavior of the system
is the same as in a bulk system.  Hence, the {\it thick film} 
regime is defined by
$\delta = \alpha q_{\perp}^2 d /B \gg 1$

The dynamics of the layers in the thick film regime is similar to that
of a bulk smectic-A.  For fixed $q_{\perp}$, when $q_z^{(n)}$ 
is large such that 
$4\rho(B \ {q_z^{(n)}}^2+K_1q_{\perp}^4)/\eta_3^2 q_{\perp}^4 > 1$, the 
smectic layers are underdamped. This is basically the second sound of the
layers, but with a correction due to surface tension and finite 
film thickness.
When $4\rho(B \ {q_z^{(n)}}^2+K_1q_{\perp}^4)/\eta_3^2 q_{\perp}^4 \ll 1$
the motion of the layers is strongly overdamped. 
The decay rates which are slow and correspond to layer undulation are
\begin{eqnarray}
  iw_-(q_{\perp},q_z^{(n)})
&=& \frac{B {q_z^{(n)}}^2+K_1 q_{\perp}^4}{\eta_3 q_{\perp}^2},
\end{eqnarray}
while the fast decay modes now correspond to the diffusive mode for $v_z$.

    To study the dynamics of the smectic-A film in the thick film regime for
typical laboratory materials, ($B\sim 2.5\times 10^7 \ {\rm dyn/cm^2}$, 
$\alpha \sim 30 \ {\rm dyn/cm}$) we need
$q_{\perp}^2 d \gg B/\alpha \sim 10^6 \ {\rm cm^{-1}}$
($\delta \gg 1$).  
For $d \sim 10^{-3}$ cm, which can be achieved easily, 
the momentum transfer in the
$xy$-plane should be at least of the order $10^5 \ {\rm cm}^{-1}$,
which 
can also be achieved.

The crossover of the dynamics for the system is illustrated in
Fig.~4, where
the normalized dynamic correlation functions
$C(q_{\perp},n,t)=c(q_{\perp},n,t)/c(q_{\perp},0,0)$
for normal modes with $n=$0, 1, 2, 3, are shown
for a smectic-A film with thickness $d=10 \ \mu$m, and typical 
elastic coefficients 
and viscosities.  The in-plane wave vector $q_{\perp}$ is chosen 
such that the
dimensionless parameter $\delta ^{-1}=B/\alpha q_{\perp}^2 d$ is 
20000, 2000,1, and 0.1, respectively.
When $\delta ^{-1} =20000$, the normal modes with $n > 0$ are 
negligible and $n=0$ normal modes
are underdamped.  For $\delta ^{-1}=2000$ the $n> 0$ normal modes 
are still negligible but
the $n=0$ normal modes are overdamped with 
a decay rate which is not a single exponential. 
This occurs when the two time scales are close to each other,
as illustrated in Fig.~3.    For $\delta=1$ the normal modes with 
$n>0$ can be observed;
this is the crossover regime between the {\it thin film} 
and {\it thick film} limits.  For
$\delta ^{-1}=0.1$ the contribution from bulk elasticity is 
sufficiently large that the dynamic correlation functions for 
normal modes with $n>0$ are easily seen in the figure.
Similar to the undulation mode in bulk systems,  all normal 
modes decay exponentially.  This is the overdamped limit of the 
{\it thick film} regime.
The actual $q_{\perp}$ values in the figure change over many orders 
of magnitude.  Thus,
experimentally probing this entire figure requires different 
scattering probes.
    
\section{Concluding Remarks}
\label{sec:conclusion}    
The layer dynamics of a freely standing smectic-A film have 
been determined
by analyzing the response functions for the surface
displacement.  By determining the positions of the poles of the
response functions in the complex $\omega$-plane, we find not only the
frequencies but also the spatial configurations for the hydrodynamic
modes of the system.  The fact that $u$ and $v_z$ are two
non-separable hydrodynamic variables, in general, is reflected in the
form of the dynamic correlation function for $u$.

When the dimensionless thickness $\delta =\alpha q_{\perp}^2 d/B \ll 1$, 
the internal structure of the
layers is not important for the system.  For fixed $q_{\perp}$ one
pair of normal modes dominates the statics and dynamics of the system.
Furthermore, the layer fluctuations and the characteristic frequencies
for these important normal modes show basically no dependence on the
elasticity of the smectic layers, {\it i.e.}, only surface tension and
viscosity are important.  To find information on the internal
structure of the system, {\it i.e.}, the smectic layers, it is
necessary to excite other normal modes.  As discussed in
reference \cite{HY-D}, exciting the $n=1$ modes provides
information on the crossover behavior of the system from a
quasi-two-dimensional system to a three-dimensional system.  This can
be done by adjusting the experimental parameters such that 
$\delta = \alpha q_{\perp}^2 d/B$ is of order unity.  
The mode structure for 
the system
in the regime $\delta  \gg 1$ is similar to that of bulk
systems.  However, the characteristic time scales for the hydrodynamic
modes now acquire surface and finite thickness corrections as well.

To summarize, we have explored the dynamical properties of a
freely-standing smectic-A film in the linear regime. It is shown that
existing experiments were performed in the limit where only the
two-dimensional character of the film can be detected. Future
experiments on the dynamical correlations of the smectic layers can
reveal the crossover behavior of the system to a regime in which layer
elasticity begins to play a role.  Our work also provides the basis
for a future theoretical study on the dynamics of a freely-standing
smectic-A film far from equilibrium, where the system has been studied
experimentally and shows behavior which is drastically different from
that of a bulk system \cite{Wu}.  The effect of finite thickness and
surface tension will have to play a central role in a theory which
describes such dynamical behavior.

\section*{Acknowledgment}
  We thank Professor X-l. Wu for helpful discussions.  
H-Y.C is grateful for fellowship support from the University 
of Pittsburgh.
DJ is grateful for the support of the NSF under DMR9217935.

\appendix
\section{}
   We discuss permeation processes in a free-standing smectic-A 
film in this 
Appendix.  We look for solution of the equations of motion with 
boundary conditions
and external forces given in Section~\ref{sec:bulk}.
In the long wavelength, low frequency limit \cite{l-l limit} 
the solution is of the form
\begin{eqnarray}
   u = u_{el} + u_p \ ,
\end{eqnarray}
where 
\begin{eqnarray}
   u_{el}({\bf r},t) 
&=& \mbox{ } u_S({\bf q}_{\perp},\omega) \ 
   {\rm cos}\left[q_z(q_{\perp},\omega) z\right]    
  \times e^{i{\bf q_{\perp}\cdot r_{\perp}} -i\omega t
     }  \nonumber  \\
& & + u_A({\bf q}_{\perp},\omega) \ 
 {\rm sin}\left[q_z(q_{\perp},\omega) z\right]       
  \times e^{i{\bf q_{\perp}\cdot r_{\perp}}
  -i\omega t }
\end{eqnarray}               
is the approximate solution used in the text. 
It is independent of permeation constant.
The second contribution
\begin{eqnarray}
   u_p({\bf r},t) &=&  \mbox{ \ } \left[ u_p^+({\bf q}_{\perp}) \     
   		\times e^{+ (1+i) p(q_{\perp}) (z- d/2)} \
  	    	 + c.c \right]  
   	         \times e^{i{\bf q_{\perp}\cdot
                  r_{\perp}} -i\omega t} \nonumber  \\	  	
   	          & &  +\left[ u_p^-({\bf q}_{\perp}) \     
   		       \times e^{- (1+i) p(q_{\perp}) (z+ d/2)} \
   		        + c.c \right]			
  	          \times e^{i{\bf q_{\perp}\cdot r_{\perp}} -i\omega t} \ , 
\end{eqnarray}  	          		
contains an inverse length, 
$p(q_{\perp}) = \sqrt{{q_{\perp}}/{2 (\zeta _P \eta _3)^{1/2}} }$,
which characterizes the decay of permeation away from the free surfaces.
In typical materials  $p(q_{\perp}) \gg q_{\perp}
           \gg q_z$ in the low frequency, long wavelength
           regime \cite{l-l limit}.
The amplitudes $u_p^{\pm}$ contain the contributions from permeation
and are related to $u_{el}$
through the full boundary conditions.

From Eqs.~(\ref{dvdz}) and (\ref{dudz}) we find that 
$|u_p^{\pm}| \ll |u_{el}|$, {\it i.e.}, 
the contribution of permeation to the layer displacement 
is small~\cite{HY-D}.  
Then Eq.~(\ref{zeta-u}) leads to 
\begin{eqnarray}
  \zeta^{\pm} \approx [u_{el}]_{z= \pm d/2} \ .
\end{eqnarray}  

It is straightforward to show that the elastic
free energy is 
\begin{eqnarray}
 F&=&\frac{1}{2} \ \int dV 
        \left\{B \left(\frac{\partial u}{\partial z}\right)^2 + 
        K_1 \left(\frac{\partial ^2 u}{\partial x^2} 
        + \frac{\partial ^2 u}{\partial y^2}\right)^2 \right\} \nonumber \\ 
 & &  +\sum_{i=+,-} \frac{1}{2} \ \int dS^{i} \   
        \alpha \left\{ 
                  \left(\frac{\partial \zeta^{i}}{\partial x}\right)^2 +
                  \left(\frac{\partial \zeta^{i}}{\partial y} \right)^2
               \right\} \nonumber  \\ 
  &\approx&\frac{1}{2} \ \int dV 
        \left\{B \left(\frac{\partial u_p}{\partial z}\right)^2 + 
               B \left(\frac{\partial u_{el}}{\partial z}\right)^2 +
        K_1 \left(\frac{\partial ^2 u_{el}}{\partial x^2} 
        + \frac{\partial ^2 u_{el}}{\partial y^2}\right)^2 \right\} 
        \nonumber \\
  & &+\sum_{i=+,-} \frac{1}{2} \ \int dS^{i} \   
        \alpha \left\{ 
                  \left(\frac{\partial u_{el}}{\partial x}\right)^2 +
                  \left(\frac{\partial u_{el}}{\partial y} \right)^2
               \right\}_{z=\pm d/2} \ .     
\end{eqnarray}   
In the linear theory the contribution from permeation can be separated
from the contribution from $u_{el}$.  Since $|u_p^{\pm}| \ll |u_{el}|$
in the regime where we perform this long wavelength analysis, we can
neglect the $u_p$ part.  However, it is $u_{el}$, not the full
displacement $u$, that can be expanded as a linear combination of the
normal modes discussed in the text and satisfies bc 2'
~(Eq.(\ref{nmode})).  When $u_p$ is included, the permeation force $B
\partial u/ \partial z$ vanishes on the free surfaces due to the
existence of $u_p$.  This is noted in the literature \cite{Kleman} but
has not been included in the previous study of the static
\cite{Holyst} and dynamic \cite{Toner} theory for free standing
smectic-A films.

\section{}
    In this Appendix we calculate the dynamic correlation functions 
for $u$ and $v_z$.
We consider the $u_{el}$ part only; the $u_p$ part simply 
corresponds to the 
permeation mode discussed in the literature \cite{DeGennes1,Martin}.  
Our starting
point is the equations of motion, {\it i.e.}, Eqs.~(\ref{NS}) 
and (\ref{dudt}),
and the dynamics of the normal modes labeled by
$({\bf q}_{\perp}, n)$ are considered.  
Eliminating the pressure via the incompressibility condition, we
find that there are two diffusive modes associated with $v_x$ and 
$v_y$, but
$v_x$ and $v_y$ decouple from $v_z$ and $u$ \cite{Lubensky}.  
Neglecting permeation,~\cite{permeation}
the equations for $v_z$ and $u$ can be expressed as
\begin{eqnarray}
  \frac{\partial}{\partial t} \  \left( \begin{array}{c} u \\ v_z 
          \end{array} \right)
 = \left( \begin{array}{cc} 0 & 1 \\ -R & -D \end{array} \right) \ 
  \left( \begin{array}{c} u \\ v_z \end{array} \right)
\end{eqnarray}
where   
\begin{eqnarray}
   D = \frac{1}{\rho} \ \eta_3 q_{\perp}^2  \ ,
\end{eqnarray}
represents the dissipative part, and 
\begin{eqnarray}
   R = \frac{1}{\rho} \ \left( B {q_z^{(n)}}^2 + K_1 q_{\perp}^4 \right)  
\end{eqnarray}
represents the reactive part.  We have used the fact that 
$q_z^{(n)} \ll q_{\perp}$ for all $n$ in the regime of our calculation
to simplify the expressions.  
The eigenvectors of the above matrix equation are
\begin{eqnarray}
   \Phi _{\pm} = \frac{1}{\sqrt{1+(i\omega_{\pm}(q_{\perp},q_z^{(n)}))^2}} \ 
               \left( 
               \begin{array}
                     {c} 1 \\ -i \omega_{\pm}(q_{\perp},q_z^{(n)})  
               \end{array}
                     \right) \ , 
\end{eqnarray}
with eigenvalues $-i \omega_{\pm}(q_{\perp},q_z^{(n)})$ 
given in Eq.~(\ref{iomega}).
Straightforward algebra leads to the following relation
\begin{eqnarray}
 \left( \begin{array}{c} u(t) \\
                         v_z(t)  \end{array}   \right)
= \frac{1}{i\omega_+ -i\omega_-} \
   \left( \begin{array}{c}
           [ i\omega_+ e^{-i\omega_- t} - i\omega_- e^{-i\omega_+ t} ] u(0)
          +[ e^{-i\omega_- t} - e^{-i\omega_+ t} ] v_z(0)  \\
          -\omega_+ \omega_- [ e^{-i\omega_+ t} - e^{-i\omega_- t} ] u(0)
          +[ i\omega_+ e^{-i\omega_+ t} - i\omega_- e^{-i\omega_- t} ] 
             v_z(0)  \end{array}
          \right)
\end{eqnarray}
where we used simplified notation, 
$\omega_{\pm} = \omega_{\pm}(q_{\perp},q_z^{(n)})$, 
$u(t)= u(q_{\perp},n, t)$, $v_z(t)= v_z(q_{\perp},n, t)$, etc.
From this one concludes that the dynamic correlation function for $u$ is
\begin{eqnarray}
& & < u(q_{\perp},n, t) \  u(-q_{\perp},m, 0) > \nonumber \\ 
&=& \delta_{nm}  \
    \frac{i\omega_+ e^{-i\omega_- t} - i\omega_- e^{-i\omega_+ t}}
     {i\omega_+-i\omega_-}
    \times
    < |u(q_{\perp},n, 0)|^2 >   \nonumber                        
    \\
&=& \delta_{nm}  \
   \frac{i\omega_+ e^{-i\omega_- t} - i\omega_- e^{-i\omega_+ t}}
{i\omega_+-i\omega_-} 
   \times
   \frac{k_B T}{B {q_z^{(n)}}^2 + K_1 q_{\perp}^4} \  ,
\end{eqnarray}                          
as given in Section~\ref{sec:film}.  The calculation for the correlation
function for $v_z$ is similar to that for $u$.

\newpage

{\large \bf  Figure Captions}

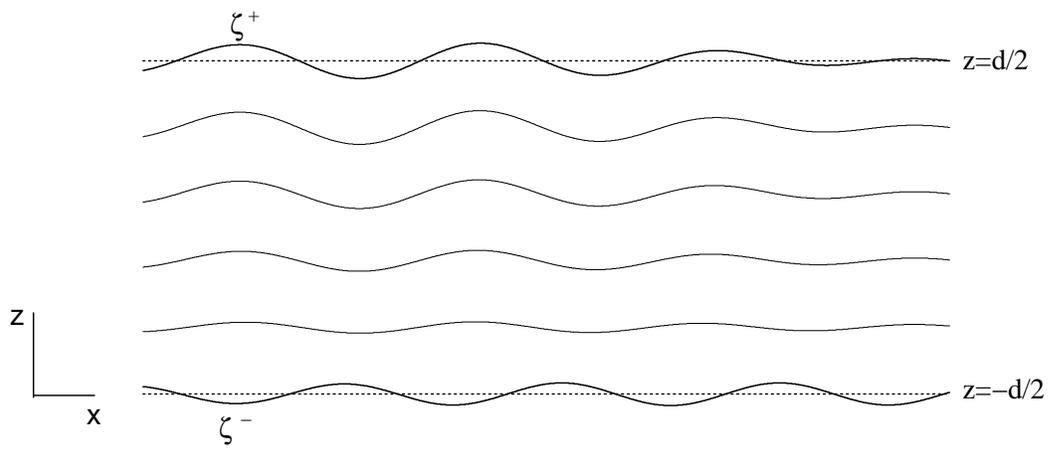
 Figure 1.
      Schematic of a freely standing smectic-A film of thickness d. 
      The $y-$axis is pointing into the paper.  The dotted lines are the
      equilibrium positions of the free surfaces.

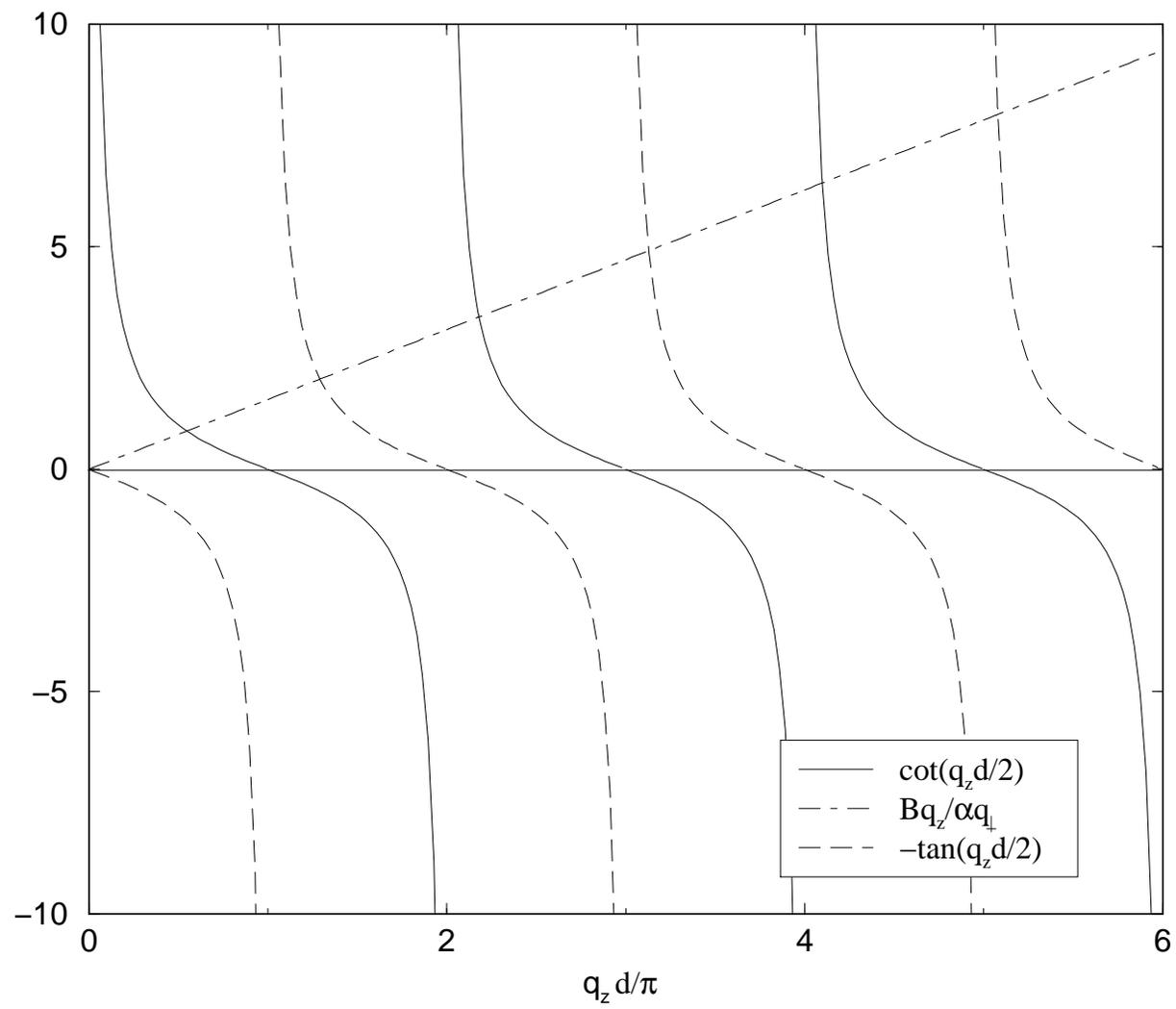
 Figure 2.
      Graphical solution for the $q_z$'s for dimensionless
 parameter $B/\alpha q_{\perp}^2d = 0.5$. The abscissa in all cases
 is $q_z d/\pi$. 
 The three functions plotted are:  $B q_z/\alpha q_{\perp}^2$ 
(dot-dashed line), 
$\cot(q_z d/2)$  (solid lines), and $-\tan(q_z d/2)$ (dashed lines).
 The crossings of the dot-dashed line and the solid lines give the
 values of is 
 $q^{(n)}_{z} d/\pi$ for normal modes with $n$ even;
 the crossings of the dot-dashed line and the dashed lines
 (except the origin) give the values of
 $q^{(n)}_{z} d/\pi$ for normal modes with $n$ odd. 
 
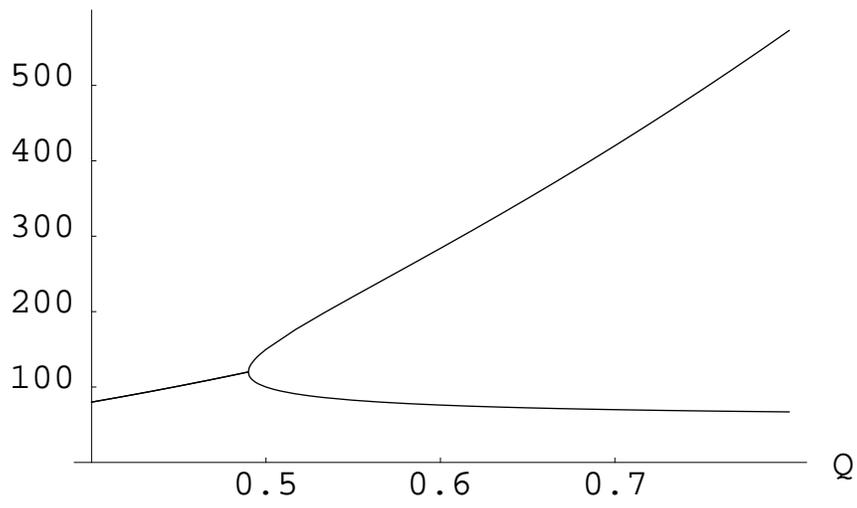
 Figure 3. 
      Decay rates for normal modes with $n=0$ in the thin film regime. 
 The dimensionless wave vector is defined as $Q=q_{\perp} d$.  The
 material parameters are chosen as $\eta _3=1.0$ p, 
 $\rho = 1.0 \ {\rm g/cm}^3$,
 $d=10 \ \mu$m, and $\alpha = 30 $ dyn/cm.  When the decay rates merge, 
 the system undergoes underdamped motion.

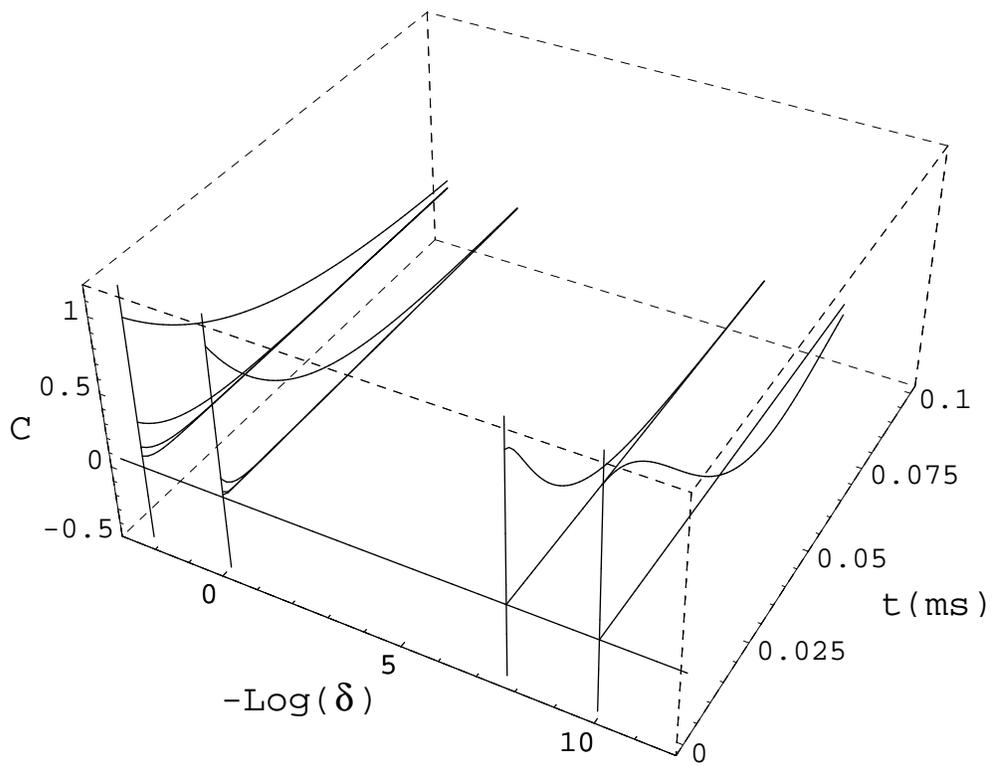
 Figure 4.
      Dimensionless dynamic correlation function 
      $C(q_{\perp},n,t) \equiv c(q_{\perp},n,t)/c(q_{\perp},0,0)$,
for $n=0,1,2,3$.  The material parameters are chosen to be
$d=10 \ \mu {\rm m}$, $B = 2.5 \times 10^7 \ {\rm dyn/cm^2}$, 
$K_1 = 10^{-6}$ dyn, $\alpha = 30 \ {\rm  dyn/cm}$, 
$\rho = 1.0 \ {\rm g/cm}^3$, $\eta _3=1.0$ poise, 
and $q_{\perp}$ is
chosen such that  $\delta ^{-1}=B/\alpha q_{\perp}^2 d  = 20000,$
2000, 1, and 0.1, respectively.    


\begin{references}
\bibitem{l-l limit} We consider the regime where $\lambda q_{\perp} \ll 1$,
                    $\zeta_p \eta_3 q_{\perp}^2 \ll 1$ (small boundary layer
                    \cite{DeGennes1}).
                    Also we assume $|\omega| \eta_3/B
                    \ll 1$, and $|\omega|/q_{\perp} \ll \sqrt{B/\rho}$,
                    the latter inequality stating that the velocity 
                    of the surface wave is small compared to the typical
                    velocity of `` second sound.''
\bibitem{com.HY-D}  To make a connection with a bulk smectic-A system,
                    the notation we choose in this article is different
                    from reference~\cite{HY-D}; however, the calculation
                    is along the same lines.
\bibitem{permeation} Permeation has negligible effect on the mode structure of
                     the $u_{el}$ part.
\end{references}
\end{document}